\documentclass[aps,onecolumn,10pt,final, showpacs]{article}
\usepackage{amsmath,amssymb,amsfonts}
\usepackage{graphics,graphicx}
\usepackage[unicode,colorlinks,linkcolor=blue]{hyperref}
\usepackage[cp1251]{inputenc}
\usepackage{color}
\usepackage{verbatim}
\usepackage{latexsym}
\begin{document}

\title{Higgs-like field and extra dimensions}

\author{A.V. Grobov\thanks{alexey.grobov@gmail.com} \and S.~G.~Rubin\thanks{sergeirubin@list.ru} \\National Research Nuclear University ``MEPhI''}

\date{}
\maketitle

\abstract{We study the origin of the Higgs field in the framework of the universal extra dimensions. It is shown that a Higgs-like Lagrangian can be extracted from a metric of an extra space. The way to distinguish our model and the Standard Model is discussed.}

\section{Introduction}

It is known that masses of particles can be associated with a hypothetical scalar field called the Higgs field. Particles of the Standard Model acquire their masses through interactions with the Higgs field. There is some set of nontrivial problems concerning the Higgs field \cite{Higgs0,Higgs1,Higgs2,Higgs3,Higgs4} and its origin is one of them. Recently Large Hadron Collider (LHC) collaboration has succeeded in searching some Higgs-like particle but its properties are still uncertain and could differ from that of the Standard Model. This has led to a growing interest in studying of various ways of the Higgs field origination. Let us briefly review some of the well-known Higgs models.

The Higgs mechanism in the framework of the Standard Model (SM) \cite{1}, \cite{2}, \cite{3} is realized by the complex scalar field that is transformed under the fundamental representation of the $SU(2)\times U(1)$ group. The potential is chosen in the form of Mexican hat which is the simplest form allowing symmetry breaking and a nonzero vacuum value of a field.
The vacuum expectation value of the Higgs field can be found from the experimental data of $\mu$-decay
\begin{equation}
v = (\sqrt{2}G_F)^{-\frac{1}{2}} = 246\ GeV
\end{equation}
where $G_F$ - Fermi constant.

As was mentioned above the SM suffers some intrinsic problems (mass hierarchy, origin of SSB and so on). This indicates that the SM is a low-energy limit of a more general theory. The idea of supersymmetry (SUSY) \cite{4}, \cite{5} provides a good basis for such a theory. Supersymmetry stabilizes the mass hierarchy of the gauge bosons and eliminates the quadratic radiative corrections.

The theory of strong dynamics (technicolor theory (TC) \cite{6}, \cite{7}, \cite{8}) gives a fresh insight into the problem of the spontaneous electroweak symmetry breaking. By analogy with the QCD, the TC leads to the confinement phenomenon, formation of a fermion chiral condensate and also has a mass scale of the order of the weak scale. The physical spectrum of the TC-theory consists of the TC singlets: techibaryons and technimesons. Complete theory should describe techniquark decays into light leptons and quarks since no stable TC objects are found. Such a theory should also provide masses of quarks and leptons.
Technicolor theory refers to the so-called higgsless theories. After the discovery of a new particle - candidate for the Higgs boson, made at July, 2012 the destiny of this theory is vague.

In the Little Higgs Model \cite{9}, \cite{10}, pseudo-Nambu-Goldstone field plays the role of the Higgs field. Due to unbroken global symmetries, the mass of the Higgs boson does not contain the one-loop quadratic divergences. New fields must be introduced to ensure that the global symmetries are not broken too strongly. At the same time these fields should cut off the quadratically divergent top, gauge, and Higgs loops.
From the beginning, the model contains the Higgs field in a form of a massless Nambu-Goldstone field. Gauge interactions, Yukawa couplings and quadratic terms of the potential are absent. The inclusion of gauge interactions into the model allows the Nambu-Goldstone boson (little Higgs) to acquire mass through loop corrections. More information about the little Higgs models can be found in the Little Higgs review \cite{LittleHiggs}.

This paper presents one of the realizations of the Higgs mechanism. Our approach is based on the
idea of the universal extra dimensions (UED), developed in the
pioneeric papers \cite{UED1}, \cite{UED2}, \cite{UED3}, \cite{UED4},
and now one of the main directions of the multidimensional gravity.

Extra components of the multidimensional metric tensor generate scalar and vector fields due to the reduction to the four dimensional space-time. The Higgs boson is described by a set of specific off-diagonal components of the metric tensor and possesses the ordinary features in the low-energy limit.

Our study follows the general direction associated with the UED  - \cite{UED5}, \cite{UED6}, \cite{Ru1}, \cite{Ru2} and based on the idea elaborated in the paper \cite{BOBRU}. In the latter paper, the form of Higgs-like potential depends on initial conditions of extra metric evolution what leads to problems with extracting specific values of the initial parameters from an experimental data. In the present paper we obtain the strict connection between the Lagrangian parameters and initial parameters like a radius of the extra space.

This approach allows not only to reproduce the conventional form of the Higgs sector but also to find deflections what became topical nowadays. It is shown that a comparison of the self couplings of the Higgs-like field to that of the Standard Higgs could give significant information provided that the LHC luminosity will be increased.

\section{Proto-Higgs field and Higgs-like potential}

According to the previous discussion, the first step consists of extracting a Higgs-like field from a metric tensor in the framework of multidimensional gravity. The Higgs field is associated with some metric components of extra space that are transformed according to the fundamental representation of the $SU(2)\times U(1)$ group.

To be more concrete, consider a $D=10$-dimensional Riemannian manifold $V_{10}=M_4 \times V_4 \times V_2$ with the metric tensor of the form:
\begin{eqnarray}\label{totalmetric}
    G_{AB} = \left(
  \begin{array}{c|c|c|c}
        g_{\mu\nu}(x) & & &     \\ \hline
        & G^{(4)}_{ab} &  e_{a}(x)   \\ \hline
                     & e_{b}(x) &-r_{d}^2 \\ \hline
                     & & &-r_{d}^2\sin^2\theta .
  \end{array}\right)
\end{eqnarray}
Here $\mu, \nu =1,2,3,4$. The subspace $V_4$ with metric
$G^{(4)}_{ab}=-r^2 _c diag(1,1,1,1)$ is described by coordinates
$y_a, a =5,6,7,8$. Our 4-dim space is described by the coordinates $x$ and a metric tensor $g_{\mu \nu}$. The second subspace $V_2$ has the geometry of sphere with radius $r_d$.

The components $e_{a}\equiv g_{a9}$ transform under fundamental representation of a linear group of coordinate transformations in $V_4$.
It will be shown below that this property allows one to connect them to the Higgs field. The metric components which are important for the following consideration are written explicitly in expression \eqref{totalmetric}.

Let us start with a non-linear $F(R)$ theory in D-dim space. The model is specified by the action
\begin{equation}\label{Act0}
S=\frac{m^{D-2} _D}{2}\int d^D X\sqrt{G}\cdot F(R)
\end{equation}
where $m_{D}$ is the D-dimensional Planck mass and $F(R)$ is an arbitrary function of the Ricci scalar.

After standard calculations the scalar curvature $R$ of the space $V_{10}$ acquires the form
\begin{eqnarray}\label{Ricciscalar}
&&R=R_{4}+K(e^2)(\partial_\mu e_{a})^2 +R_6 (e^2), \\
&&K(e^2)=\frac{1}{2}\frac{r^2 _c r^2 _d-5e^2}{(e^{2} - r^2 _c r^2 _d)^2} \label{h} \\
&&R_6(e^2)=\frac{2r^2 _c}{ r^2 _c r^2 _d - e^2} \label{V}
\end{eqnarray}
where metric (\ref{totalmetric}) is taken into account. We also use the notations $e^2 =e_a^2 = e_5 ^2 + e_6 ^2 +e_7 ^2 +e_8 ^2$ valid for the Euclidean geometry of the subspace $V_{4}$  and $e =\sqrt{e^2}$.

To facilitate the analysis, slow motion approximation \cite{BOBRU} will be used. In our case this approximation is correct if the Ricci scalar $R_6$ of the extra space is large comparing to the other terms in \eqref{Ricciscalar}. The condition of slow motion
\begin{equation}\label{R6}
R_6 (e^2) \gg \alpha ; \quad \alpha \equiv R_{4}+K(e^2)(\partial_\mu e_{a})^2
\end{equation}
allows one to use first terms of the Tailor series
\begin{equation}\label{FR}
F(R)=F(R_6 + \alpha)=F(R_6)+\alpha \cdot F'(R_6).
\end{equation}
Numerical values of the parameters used below lead to the Ricci scalar
$$R_6 (e^2)\sim 10\, TeV^2.$$
and one can easily check that inequality (\ref{R6}) holds true at temperatures much smaller than $10$ TeV.

Action \eqref{Act0} with Ricci scalar \eqref{Ricciscalar} can be integrated over the extra dimensional variables. This leads to the effective 4-dimensional action
\begin{equation}\label{Act1}
    S=\int d^{4}x \sqrt{-g} \left[\varphi(e^2)\frac{M^2}{2}R_{4}+\psi(e^2)(\partial_\mu e_{a})^2-U(e^2)\right]
\end{equation}
where
\begin{equation}\label{m4}
 \frac{M^2}{2} = (2\pi)^{5}m^8 _{D}r^4 _c r^2 _d
\end{equation}
\begin{equation}\label{phi} \nonumber
\varphi(e^2) =f(e^2)F'(R_6)
\end{equation}
\begin{equation}\label{psi}\nonumber
\psi(e^2) =\frac{M^2}{2}\varphi(e^2)K(e^2)
\end{equation}
\begin{equation}\label{U}\nonumber
U(e^2) = -\frac{M^2}{2}f(e^2)F(R_6)
\end{equation}
\begin{equation}\label{f(e)}\nonumber
    f(e^2)=\sqrt{\left|1 -\frac{e^{2}}{r^{2} _c r^{2} _d}\right|}
\end{equation}

After the conformal mapping, see e.g. \cite{BronnikovMelnikov}
\begin{eqnarray}\label{Conftransform}
    g_{\mu\nu}&=&\Omega(e^2)\overline{g}_{\mu\nu} \\
     \Omega(e) &=& |\varphi(e^2)|^{-\frac{2}{D-2}} \nonumber
\end{eqnarray}
and the change of variables $e_a \rightarrow h_a$
$$ h_a = \int \frac{\sqrt{2|L(e_a ^2)|}}{\varphi(e_a ^2)} de_a ,$$
the action acquires the Einstein-Hilbert form in terms of the scalar field $h_a$.
\begin{equation}\label{Confaction}
   S=\int d^{4}x \sqrt{-\overline{g}} \left[sign(\varphi)\left[\frac{M^2}{2}\overline{R}_{4}+sign(L)\frac{1}{2}(\partial_\mu h_{a})^2 \right] - V(h^2)
   \right] .
\end{equation}
Here the notations
\begin{equation} \nonumber
L(e_a ^2) = \varphi(e^2)\psi(e^2)+\frac{3}{2}\frac{M^2}{2}\left(\frac{\partial \varphi(e^2)}{\partial e}\right)^2
\end{equation}
and
\begin{equation}\label{Confpotential}
   V(h^2)=|\varphi(e^2)|^{-2}U(e^2),
\end{equation}
were introduced. The effective potential $V$ depends on the new field $h_a$ as long as $e_a=e_a (h)$. Equalities $sign(\varphi)=sign(L)=1$ are checked to be true in our case.

The standard form of the Hilbert-Einstein action with the $\Lambda$ term is
\begin{equation} \nonumber
S = \int d^{4}x \sqrt{-g} \left[ \frac{M^2 _{pl}}{2} R - 2\Lambda \right]
\end{equation}
In our case the cosmological constant is hidden in the potential $V(h^2)$.
We assume that the effect we study here is not the only one that contributes to the cosmological constant and the Planck mass. In particular one could add an additional extra space so that its Ricci scalar contributes to the vacuum energy and shifts the cosmological term to the observed value. Moreover the values of these parameters are not important for particle physics considered in the Minkowski space. In this case the Ricci scalar equals zero and we may omit the first term in \eqref{Confaction}.

Till now we did not discuss the form of $F(R)$ function to maintain the general form of equations and formulas.
As a next step let us choose specific form of the function $F(R)$
\begin{equation}\label{F}
F(R)=R+cR^2+bR^3-2\Lambda .
\end{equation}

The following analysis is strongly simplified if the metric fluctuations $e$ are small:
\begin{equation}\label{simplification}
    e^{2} _a << r^{2} _c r^{2} _d .
\end{equation}
This checked to be true in the vicinity of the minimum of the potential, where
$
e^{2}_a / (r^{2} _c r^{2} _d) \approx 0.009 << 1
$
for the parameter values $r_c=r_d=0.06$, $m_D = 3.29$ and $\eta=246$ GeV, see the figure caption. Moreover the field $h_a$ is proportional to the field $e_a$ with two percent accuracy,
\begin{eqnarray}\label{heconnection2}
     h_a \simeq \frac{M}{\sqrt{2}r_c r_d}e_a,
\end{eqnarray}
in this case.
This strongly facilitates the analysis and leads to the explicit form of potential \eqref{Confpotential}
\begin{eqnarray}\label{Confpotential3}
   V(h_a^2)&=&\frac{M^2}{2r^2 _d}\frac{1}{\sqrt{1-\frac{2 h_a ^2}{M^2}}} \left( 1+\frac{c_1}{1-\frac{2 h_a ^2}{M^2}} +\frac{c_2}{\left(1-\frac{2
   h_a ^2}{M^2}\right)^2}\right)^{-2} \\ \nonumber
   &\times&\left[ -\frac{2}{1-\frac{2 h_a ^2}{M^2}} -\frac {c_1}{\left(1-\frac{2 h_a ^2}{M^2}\right)^2} -\frac{\frac{2}{3}c_2}{\left(1-\frac{2 h_a ^2}{M^2}\right)^3} +2 \Lambda r^2 _d\right] ,
\end{eqnarray}
where $c_1 = 4 \cdot c \cdot r^{-2} _d , c_2 = 12 \cdot b \cdot r^{-4} _d$ and expression \eqref{F} is used.

Let us express the four-component column $\hat{h}=( h_5,h_6,h_7,h_8 ) $ in terms of two-component complex fields $X,Y$ in the following way
\begin{equation} \label{invtrans}
    \hat{h}\equiv\left(
          \begin{array}{c}
            X \\
            Y \\
          \end{array}
        \right)
\end{equation}
and define new two-component complex column $H$ - the Higgs field - as
\begin{equation} \label{hXY}
H=X + iY.
\end{equation}

The correspondence between the fields $H$ and $h$ can be described by using the matrix
\begin{equation}\label{P} P=\frac{1}{\sqrt{2}}\left(
    \begin{array}{cccc}
      1 & 0 & i & 0 \\
      0 & 1 & 0 & i \\
    \end{array}
  \right)
\end{equation}

  The matrix (\ref{P}) ``projects''\ one representation of the Higgs field,
  $\hat h=(h_5, h_6, h_7, h_8)$, onto the other,
 \begin{equation} \label {HPh}
     H=Ph.
 \end{equation}
  so that
\begin{equation}\label{Hh}
    H=\left(    \begin{array}{c}
            h_5 + ih_7 \\
            h_6 + ih_8 \\
      \end{array}    \right)
\end{equation}
The transformation properties of this field are discussed in the next Section.

In terms of the field $H$ the Lagrangian can be obtained by combining equations \eqref{Confpotential3} and evident equality
   $ h_a h_a = H^+ H.$
Its final form
\begin{eqnarray}\label{Higgslike}
&& L_H=\frac{1}{2}(\partial_{\mu}H)^+ (\partial_{\mu}H) - V(H^+ H),\\ \nonumber
&&V(H^+ H) = \frac{M^2}{2r^2 _d} \frac{1}{\sqrt{1-\frac{2H^+ H}{M^2}}} \left(
1+\frac{c_1}{1-\frac{2H^+ H}{M^2}} +\frac{c_2}{\left(1-\frac{2H^+
H}{M^2}\right)^2}\right)^{-2} \\ \nonumber && \times \left[-2
\left(1-\frac{2H^+ H}{M^2}\right)^{-1}-c_{1} \left(1-\frac{2H^+
H}{M^2}\right)^{-2} - \frac{2}{3}c_{2} \left(1-\frac{2H^+
H}{M^2}\right)^{-3} + 2\Lambda r^2 _d \right] \nonumber
\end{eqnarray}
 includes the same parameter values as in \eqref{Confpotential3}

\begin{figure}[h!]\label{pot1}
\includegraphics[scale=0.4]{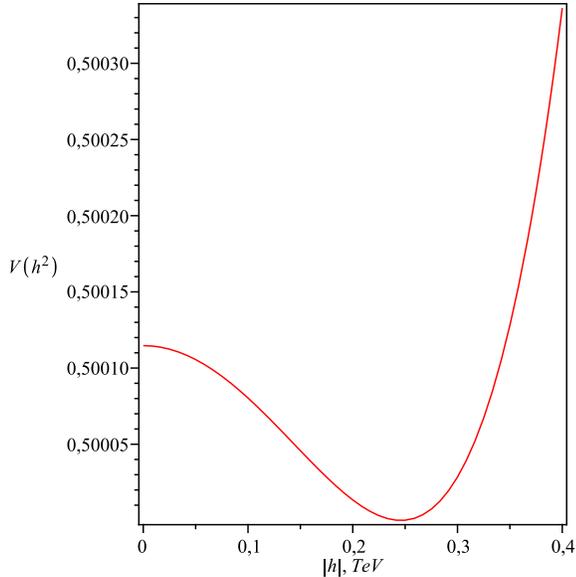}
\caption{Higgs-like potential. The parameter values are $r_c=r_d=0.06$ TeV$^{-1}$, $m_D = 3.29$ TeV, $c_1=10061.94$, $c_2 = -6244.86$, $2 \Lambda r_d^2 = 10100.03$}
\end{figure}

The form \eqref{Higgslike} does not look like the standard Higgs potential. Nevertheless, it has appropriate low-energy behavior for some parameter values which can be found from the conditions
\begin{equation} \label{cond1}
\frac{d}{dh}V(h^2 = \eta^2 ) = 0,
\end{equation}
\begin{equation}\label{cond2} \nonumber
\frac{d^2}{dh^2}V(h^2 = \eta^2 ) = m^2 _h,
\end{equation}
where $h=\sqrt{H^+ H}=\sqrt{h_5 ^2 + h_6 ^2 +h_7 ^2 +h_8 ^2} $ (see  \eqref{Hh}) and $\eta=246$ GeV is the standard Higgs vacuum expectation value. Recent results on Higgs search indicate that its mass $m_h$ is about 125 GeV and we base on this value to determine the parameters of our model. Numerical calculation of the Higgs potential \eqref{Confpotential3} of our model, is represented in Figure \ref{pot1}.

Let us compare the coupling constants derived from the Standard Model to the one derived from our model. In our model, Taylor series of the Higgs-like potential include an infinite number of vertexes
\begin{equation}\label{taylorseries2}
   V(h^2)=V(\eta) + \frac{m^2 _h}{2}\rho(x) ^2 + \lambda_1
   \rho(x) ^3 +\lambda_2 \rho(x) ^4 + \lambda_3 \rho(x) ^5+...,
\end{equation}
where $h=\eta + \rho(x)$.
Numerical calculation of the coupling constants $\lambda_1,\lambda_2$ and so on gives the following numbers
\begin{eqnarray}\label{lambdaOUR}
   \lambda_1=0.0337, \quad
   \lambda_2=0.0445, \quad
   \lambda_3=0.0279; \quad V(\eta )=0.5
\end{eqnarray}
Recall that $m_h=0.125$ TeV and $\eta =0.246$ TeV.

As was mentioned above an absolute value of the potential is not important
since gravitation has not been taken into account. Nevertheless we represented this value for completeness.

The Higgs potential in the Standard Model has the form
\begin{equation}\label{taylorseries}
   U_{SM}(H)=\frac{m^2 _h}{2}\rho(x) ^2 + \lambda_1
   \rho(x) ^3 +\lambda_2 \rho(x) ^4
\end{equation}
where the numerical values of coupling constants
\begin{eqnarray}\label{lambdaSM}
   \lambda_1&=&0.0317, \quad \lambda_2=0.0322, \quad \lambda_{n>2}=0
\end{eqnarray}
are connected to the Higgs mass which was determined recently.

One can see that the corresponding coupling constants in expressions \eqref{lambdaOUR}, \eqref{lambdaSM} have different values. This point could be checked in the forthcoming LHC experiments. Non-zero value of the coupling constant $\lambda_3$ represents crucial difference between Standard Model and our one.

\section{Interaction between the Higgs-like field and the gauge fields}

Up to now we have not considered the transformation properties of the field $h_{a}$ and its relation to the Higgs field $H$ of the Standard Model. This point was discussed in \cite{BOBRU} and we shortly remind the idea.

In our approach, symmetries of a Lagrangian depend on symmetries of a compact space in the Kaluza-Klein spirit. Consider a class $T$ of linear coordinate transformations of our extra space with metric \eqref{totalmetric}
\begin{equation}\label{Ttrans}
        y'^a =T^a_{\;b} y^b, \qquad a,b=5,...,8.
\end{equation}
It is supposed that the matrices $T=T_1 T_2$ form a Lie group of isometries of the extra space. The set of $4 \times 4$ matrices $T_1,\ T_2$ is defined as follows
\begin{equation} \label{G12}
    T_1=\left(
  \begin{array}{cc}
    I\cos\phi & -I\sin\phi \\
    I\sin\phi &  I\cos\phi \\
  \end{array} \right),
    T_2=\left(
  \begin{array}{cc}
    A & -B \\
    B &  A \\
  \end{array}   \right),
\end{equation}
where $I$ is the unit $2 \times 2$ matrix and matrices $A,B$ satisfy the conditions
\begin{equation} \label{AB}
    A^T A + B^T B=1, \qquad A^T B - B^T A=0, \qquad
     \det(A+iB) = 1.
\end{equation}
Under these conditions a real 4-parametric group $T$ is isomorphic to the group $SU(2)\times U(1)$ \cite{BOBRU}.
Transformation properties of the proto-higgs fields $e_a$ and $h_a$ are the same as in \eqref{Ttrans},
\begin{equation}\label{transe}
        h'^a =T^a_{\;b} h^b, \qquad a,b=5,...,8.
\end{equation}
and hence this field transforms under fundamental representation of the group $T$.

According to \eqref{invtrans}, \eqref{hXY} the doublet $H$ is connected to the proto-Higgs field $e_a$. As was shown in \cite{BOBRU}, the described transformation properties of the proto-Higgs field leads to proper transformations of the field $H$. The latter is transformed by the fundamental representation of the electroweak group $SU(2)\times U(1)$
\begin{equation}\label{group3}
        H'=\omega_1 \omega_2 H=(A+iB)e^{i\phi}H
\end{equation}
and hence could be considered as a candidate to Higgs boson.

  We have singled out the components of the metric tensor interpreted
  as the Higgs bosons. The way of extracting gauge fields from the extra-dimensional metric is well known. Namely, the metric (\ref{totalmetric}) is represented in a  standard form (see, e.g.,
  \cite{Montani, Blagojevic}), where the following components of the total metric
  (\ref{totalmetric}) will be of interest for us:
\begin{eqnarray}  \label{one}
    && g_{\mu\nu}(x,y) = g_{\mu\nu}(x) + g_{ab}k^{a}_{i}(y)A^i_{\mu}(x)
        k^{b}_{j}(y) A^j _{\nu}(x) , \\
   \label{two}
    && g_{\mu a}(x,y)= g_{ab}k^{a}_{i}(y)A^i_{\mu} \quad i,j = 1,2,3,4. \nonumber
\end{eqnarray}
The set of Killing vectors $k^a_i$ acts in the subspace $V^4$ with the metric $G^{(4)}_{ab}$. They satisfy the relations
\begin{equation}
k^{a}_{i}\partial_a k^{b}_{j} - k^{a}_{j}\partial_a k^{b}_{i}= f_{ij}^{l}k^{b}_{l}
\end{equation}
where $f_{ij}^{l}$ are structure constants of the Lie group $T=T_1 \cdot T_2$.
It can be shown, see e.g. \cite{Blagojevic} that the fields $A^i_{\mu}(x)$ transform exactly as gauge fields under the action of the group $T$
\begin{equation}\label{transA}
A^i_{\mu}+\delta A^i_{\mu} =A^i_{\mu}-\partial_{\mu}\varepsilon^i +f_{lk}^{i} A^l_{\mu} \varepsilon^k
\end{equation}
if the infinitesimal coordinate transformations
\begin{equation}
x^{\prime\mu}=x^{\mu},\quad y^{\prime\mu}=y^{\mu}+ k^{a}_{i}\varepsilon^i
\end{equation}
in $V^4$ take place.

The structure of the term $L_{int}(A,h)$, containing interaction
between the field $h_a$ and the gauge fields $A_\mu ^i$, can be
obtained from general considerations. The initial Lagrangian is
invariant under general coordinate transformations and therefore
under those belonging to the group $T$. This also relates to
those part of the Lagrangian containing the fields $h_a$ and
$A^{i}_\mu$ as well. On the other hand
transformations \eqref{transe} and \eqref{transA} are known as gauge
transformations and hence the fields $A$ and $h$ must enter the Lagrangian in a certain gauge-invariant
combination. The latter is well-known and has the form
\begin{eqnarray}
L_{int}(A,h) &=&  g^{\mu\nu}(D_\mu h_a)^+ (D_\nu h^a), \nonumber \\
(D_\mu h)_a &=& \left(\delta_{ab}\partial_{\mu}
        +A^{i}_{\mu}(x) t_{i,a b}\right)h_b . \nonumber
\end{eqnarray}
As was shown in \cite{BOBRU} the explicit form of the group generators $t_{i,a b}$ contains the Pauli matrices $\tau_m$. The transition \eqref{HPh} to the field $H$ leads to the standard form of interaction of the gauge and Higgs fields
\begin{equation} \label{Linth}
        L_{int}(A,H)= g^{\mu\nu}(D_\mu H)^{+}(D_\nu H),
\end{equation}
where the gauge-invariant derivative of the field $H$ has the form
$$
    D_{\mu} \equiv (\partial_{\mu} + A_{\mu}^m \tau_m + B_{\mu}I), \quad B_{\mu}\equiv A_{\mu}^0 .
$$

The expression (\ref{Linth}) represents the standard form of interaction
between the gauge fields and the Higgs field belonging to the fundamental
representation of the gauge group and corresponding to the boson sector
structure of the SM.

\section{Discussion}
In this paper we study one of the possible origin of the Higgs phenomena. It was shown that the Higgs field is constructed from the metric components of the 6-dim extra space. The parameters of the effective Lagrangian i.e. the Higgs mass and the coupling constants depend on the sizes of extra spaces.

Variation of initial parameters like the extra space sizes $r_c$ and $r_d$ allows to obtain different values of the Higgs boson mass and proper vacuum expectation value equals 246 GeV. In this paper we limit ourselves by a Higgs mass about 125 GeV. Main difference between our model and the Standard Model lies in different values of coupling constants, see expressions \eqref{lambdaOUR} and \eqref{lambdaSM} for comparison. Future measurements of the 3- and 4- vertex coupling constants will clarify the situation. If the coupling constants appear to be different this will indicate a possible existence of an extra space.

The SM is proved to be a renormalizable theory, in particular due to the form of the Higgs potential. The properties and the existence of the SM Higgs field itself are postulated from the beginning. Our aim was to derive its properties by defining it as the off-diagonal components of the metric tensor of extra space. The Higgs-like potential obtained here reveals a non-renormalizability of the effective theory what is common situation when dealing with gravity. In fact, every field theory becomes "slightly nonrenormalizable" when an interaction with gravity is involved. Main signature of the nonrenormalizable character of our Higgs-like Lagrangian is the nonzero value of the parameter $\lambda_3$ in (\ref{lambdaOUR}). Thus an experimental search for the 5-vertex self-interaction of the Higgs-like particle would be very instructive though very difficult.

There are several points that must be discussed in the
future. Firstly, the problem of stabilization of extra space is not
considered here, though the gravity with higher derivatives contains
such possibility, see \cite{Ru1}. Secondly, the coupling constants
of the Higgs field and gauge ones remain uncertain. At last, we
found those components of the
extra metric which behaves like the Higgs field. Other components of
the extra metric are not discussed.

The authors are grateful to R. Konoplich for fruitful discussions. The work of AVG was supported by the grant 14.132.21.1446 of Ministry of Education and Science of the Russian Federation. The work of SGR was partially supported by the grant 14.A18.21.0789 of Ministry of Education and Science of the Russian Federation.

\end{document}